\documentclass[12pt]{article}

\usepackage[height=8.85in,width=6.45in]{geometry}

\usepackage[utf8]{inputenc}
\usepackage[T1]{fontenc}
\usepackage{amsmath}
\usepackage{amssymb}
\usepackage{mathtools}
\numberwithin{equation}{section}
\allowdisplaybreaks

\usepackage{slashed}
\usepackage{braket}
\usepackage[usenames,dvipsnames,svgnames,table]{xcolor}
\usepackage[colorlinks,citecolor=DarkGreen,linkcolor=FireBrick,urlcolor=FireBrick,linktocpage]{hyperref}
\usepackage{cite}
\usepackage{graphicx}
\usepackage{times}
\usepackage[scaled]{couriers}

\usepackage{bm}
\usepackage{subfig}

\usepackage{tikz,pgf}
\usepackage{tikz-cd}
\usetikzlibrary{shapes}
\usetikzlibrary{calc}
\usetikzlibrary{decorations.pathmorphing}
\usetikzlibrary{decorations.pathreplacing,shapes.misc}
\usetikzlibrary{positioning}
\usetikzlibrary{decorations.markings}
\tikzset{->-/.style={decoration={
  markings,
  mark=at position .5 with {\arrow{>}}},postaction={decorate}}}
\tikzset{-<-/.style={decoration={
  markings,
  mark=at position .5 with {\arrow{<}}},postaction={decorate}}}

\usepackage{xcolor}
  \definecolor{rblue}{RGB}{81, 49, 193}
  \definecolor{rorange}{RGB}{255, 147, 40}
  \definecolor{rgreen}{RGB}{176, 233, 0}

\renewcommand{\tilde}{\widetilde}

\DeclareMathOperator{\tr}{tr}
\DeclareMathOperator{\Tr}{Tr}

\DeclareMathOperator{\Hom}{Hom}
\newcommand{\spin}{\text{spin}}

\renewcommand{\hat}{\widehat}

\usepackage{mdframed}

\def\Hom{\mathop{\mathrm{Hom}}}
\def\Sq{\mathop{\mathrm{Sq}}\nolimits}
\def\bZ{\mathbb{Z}}
\def\U{\mathrm{U}}
\def\Sp{\mathrm{Sp}}

\begin{document}

\begin{titlepage}

\begin{flushright}
IPMU-19-0064
\end{flushright}

\vskip 3cm

\begin{center}

{\large \bfseries On gapped boundaries for SPT phases beyond group cohomology}

\bigskip
\bigskip
\bigskip

Ryohei Kobayashi$^1$, Kantaro Ohmori$^2$, and Yuji Tachikawa$^3$
\bigskip
\bigskip
\bigskip

\begin{tabular}{ll}
 $^1$ & Institute for Solid State Physics, \\
 &University of Tokyo, Kashiwa, Chiba 277-8583, Japan\\
 $^2$ & School of Natural Sciences, Institute for Advanced Study, \\
 & 1 Einstein Drive Princeton, NJ 08540, USA \\
$^3$ & Kavli Institute for the Physics and Mathematics of the Universe (WPI), \\
& University of Tokyo, Kashiwa, Chiba 277-8583, Japan
\end{tabular}

\vskip 1cm

\end{center}

\noindent 
We discuss a strategy to construct gapped boundaries for a large class of symmetry-protected topological phases (SPT phases) beyond group cohomology.
This is done by a generalization of the symmetry extension method previously used for cohomological SPT phases.
We find that this method allows us to construct gapped boundaries for  time-reversal-invariant bosonic SPT phases and for  fermionic Gu-Wen SPT phases for arbitrary finite internal symmetry groups.

\end{titlepage}

\setcounter{tocdepth}{2}
\tableofcontents

\section{Introduction and summary}
An important feature of  a topological phase of matter is that it often supports a nontrivial theory on its boundary.
For example, in the case of the 2+1d quantum Hall system, the boundary is a theory of gapless chiral charged fermion in 1+1 dimensions.
This boundary theory has an anomaly under the electromagnetic $\U(1)$ symmetry,
which is cancelled by the gauge variation of the bulk theory localized at the boundary.
This prototypical case has been generalized in many directions in the recent years,
and this study led us to the realization that the anomaly of a $(d-1)+1$ dimensional system with symmetry $G$ is characterized by the corresponding $d+1$ dimensional bulk topological phase, 
known under the general name of a symmetry-protected topological phases (SPT phase) protected by the symmetry $G$.\footnote{%
In this paper we do not make a careful distinction between invertible phases, SPT phases in the strict sense, and SPT phases in the general sense.
Invertible phases are low-energy limits of gapped phases with a unique ground state on an arbitrary closed spatial manifold.
SPT phases are obtained by identifying  invertible phases which can be connected by continuous deformations.
SPT phases in the strict sense and the general sense are distinguished by whether or not they become trivial when the (internal) $G$ symmetry is not imposed.
}

In this context, it is a basic question to answer which nontrivial SPT phase supports a gapped boundary.\footnote{%
Note that it is not immediate that every SPT phase admits any boundary at all in the first place.
For example, $\Omega^{\spin}_{4n}(\mathrm{pt})$ has a quotient $\simeq \bZ^{r(n)}$ related to the polynomials of Pontryagin classes, which corresponds to various spin SPT phases in $4n-1$ dimensions.
It is a nontrivial mathematical theorem \cite{ABP} that they can be detected by KO-Pontryagin classes. 
Physically, this means that these SPT phases represent anomalies of free fermions which can have additional tangent bundle indices.
}
On the one hand, there are certainly SPT phases whose boundary are forced to be gapless as discussed e.g.~in \cite{Sodemann:2016mib,Wang:2017txt,Garcia-Etxebarria:2017crf,CordovaOhmoriToAppear}.
On the other hand, there is also a large class of nontrivial SPT phases which are known to admit gapped boundaries, some of which are described in \cite{Seiberg:2016rsg,Witten:2016cio,Wang:2017loc,Tachikawa:2017gyf,Hsieh:2018ifc,Guo:2018vij,WangOhmori2018,ThorngrenKeyserlingk2015,Kobayashi2019,Kapustin2014anomalous,Fidkowski2014,Cheng2017exactly,Chen2016symmetry,Chen2015anomalous,Barkeshli2016}.

Among them, the most systematic method known thus far is based on the symmetry extension method, originally introduced in \cite{Witten:2016cio} and systematized in \cite{Wang:2017loc}.
In particular, this method is known \cite{Wang:2017loc,Tachikawa:2017gyf} to produce gapped boundaries for bosonic SPT phase for any finite internal symmetry group $G$ described by the group cohomology \cite{Chen:2011pg}.
The aim of this note is to describe how this method can be adapted to SPT phases more general than these, i.e.~to SPT phases beyond group cohomology.
Such phases  are now known to be classified by the suitable dual of the bordism group $\Omega_{d+1}^H(BG)$, where $G$ is the global symmetry group and $H$ stands for the choice of the spacetime symmetry such as fermionic parity and/or time reversal \cite{Kapustin:2014tfa,Kapustin:2014dxa,Freed:2016rqq,Yonekura:2018ufj}.
This will be explained in Sec.~\ref{sec:general}.

We have 
two applications:
The first is given in Sec.~\ref{sec:unoriented}, where we construct gapped boundaries for time-reversal invariant bosonic SPT phases for any finite internal symmetry group $G$, which are known to be characterized by $\Omega^\text{unoriented}(BG)$. 
This result follows easily from our general construction and a mathematical theorem from the late 60s which describe the relevant bordism group explicitly.
The second is given in Sec.~\ref{sec:GuWen}, where we construct gapped boundaries for a subclass of fermionic SPT phases known as the Gu-Wen phases, originally introduced in \cite{Gu:2012ib} and studied further in \cite{Gaiotto:2015zta}. 
This will be done by extending the definition of the Gu-Wen Grassmann integral from the bulk to the coupled system of the bulk and the boundary.
Again, this allows us to construct a gapped boundary for Gu-Wen phases for any finite internal symmetry group.

\section{General construction}
\label{sec:general}
\subsection{The symmetry extension method}
\label{sec:extension}
Let us first recall the symmetry extension method to construct gapped boundaries for cohomological SPT phases, described in  \cite{Witten:2016cio,Wang:2017loc,Tachikawa:2017gyf}.
Take a class $[y] \in H^{d+1}(BG,U(1))$ where $y\in Z^{d+1}(BG,U(1))$.
The corresponding bulk SPT phase has the action $\int_{N_{d+1}} y$.
Assume that there is an extension 
\begin{equation}
0\to K\to H \stackrel{p}{\to} G \to 0 \label{extension}
\end{equation}
and the corresponding fibration 
\begin{equation}
BK\to BH \stackrel{p}{\to} BG \label{extensionB}
\end{equation}
such that  $\delta x=p^*y$ for $x\in C^d(BH,U(1))$. We let $p$ denote both of the projection between the groups $H\to G$ and the projection between their classifying spaces $BH\to BG$.

Consider the boundary gauge theory  whose partition function has the form\begin{equation}
Z \propto \sum_{h\in [M, BH] } \exp({-2\pi i \int_M h^*x}) \label{foo}
\end{equation} where we sum over the $H$ gauge fields specified by  $h\in [M, BH]$
which lifts a given $G$ background gauge field  specified by $g\in [M, BG]$.
This provides a gapped boundary with the anomaly $y$.

To see this, we explicitly show that the theory on $M$ couples to the bulk $N$ where $\partial N=M$. 
Suppose we are given $g:N\to BG$ such that it lifts to $h:M\to BH$ on the boundary. 
Then $g^*y=\delta (h^*x)$ on the boundary. 
Therefore $\int_N g^*y -\int_M h^*x$ is well-defined.

In particular,  as shown in \cite{Tachikawa:2017gyf} we can choose a $K$ with a nontrivial $G$ action such that every $y\in Z^{d+1}(BG,U(1))$ can be written as $y=e\cup z$ where $e\in Z^2(BG,K)$ is the extension class and $z\in Z^{d-1}(BG,\hat K)$.
Almost tautologically there is $a\in C^1(BH,K)$ such that $\delta a=e$,
implying $x=-a\cup z$ satisfies $y=\delta x$.

Now note that an $h:M\to BH$ lifting $g:M\to BG$ provides
 $\underline{a}:=h^*a\in C^1(M,K)$ such that $\delta\underline{a}=g^*e$.
Then the boundary gauge theory \eqref{foo} becomes  \begin{equation}
\sum_{\delta a=g^*e} \exp(2\pi i\int_M a\cup g^*z)  
= \sum_{\substack{a\in C^1(M,K),\\ b\in C^{d-2}(M,\hat K)}}
 \exp(2\pi i\int_M \left(a \cup \delta b + a\cup g^*z + (g^*e)\cup b\right))\label{bar}
\end{equation}  where $\underline{a}$ was simply denoted by $a$.

\subsection{Extension by higher-form symmetries}
We note that the action \eqref{bar} is an example of the topological Green-Schwarz mechanism.
More generally, we can consider a cochain field theory whose partition function is of the form \begin{equation}
Z\propto \sum_{\substack{a\in C^p(M,K),\\ b\in C^{q}(M,\hat K)}}
\exp({2\pi i\int_M a \cup \delta b + a\cup A + B\cup b}) 
  \propto \sum_{\substack{a\in C^p(M,K),\\ \delta a=B } }
 \exp({2\pi i\int_M  a\cup A })  \label{topGS}
\end{equation}
where $p+q=d-1$, $A\in Z^{d-p}(M,\hat K)$, $B\in Z^{d-q}(M, K)$. 
This is a $(p{-}1)$-form $K$-gauge theory\footnote{%
Our convention is that an ordinary gauge theory is a 0-form gauge theory having a one-form gauge field.
}, and couples to a $(q+1)$-form $\hat K$-symmetry background $A$
and a $(p+1)$-form $K$-symmetry background $B$.\footnote{%
For the basics of higher form symmetries, see e.g.~\cite{Gaiotto:2014kfa}.
}
This theory has an anomaly $\int_{N} B\cup A$.
Our case \eqref{bar} is when $p=1$, $A=g^*z$ and $B=g^*e$.

This means that the symmetry extension method can be generalized so that the symmetry is extended by a higher-form symmetry.
For example, say that a given $y\in Z^{d+1}(BG,U(1))$ can be written as $y=e\cup z$ where $e\in Z^{p+1}(BG,K)$ and $z\in Z^{q+1}(BG,\hat K)$.
Then the $(p{-}1)$-form $K$-gauge theory \eqref{topGS} has the anomaly $y$, by setting $A=g^*z$ and $B=g^*e$,
and the action $x:=a\cup A=a\cup g^*z$ is exactly the class $x$ which trivializes $g^*y$ via $\delta x=g^*y$.

The class $x$ itself can be considered as a pull-back via the projection of the fibration \begin{equation}
K(K,p)\to \underline{BH} \to BG\label{cf}
\end{equation} whose Postnikov class is specified by $e\in H^{p+1}(BG,K)$.
Tautologically, there is a cochain $a\in C^{p}(BH,K)$ such that $\delta a=e$ and therefore
$\delta(a\cup z)=e\cup z= y$.

The fibration \eqref{cf} is a fibration among  classifying spaces for the extension of symmetries \begin{equation}
0\to K_{[p-1]} \to \underline{H} \to G_{[0]} \to 0
\end{equation} where the subscript $[d]$ means a $d$-form symmetry,
and the underlines are used to emphasize that it represents a symmetry which mixes the ordinary 0-form symmetry and the higher $(p-1)$-form symmetry.

More generally, if the anomaly class $y\in H^{d+1}(BG,U(1))$ is trivialized in $\underline{BH}$ so that there is an $x\in C^{d}(\underline{BH},U(1))$ such that $\delta x=y$,
we can simply consider the $d$-dimensional $(p{-}1)$-form $K$-gauge theory whose partition function is \begin{equation}
Z\propto \sum_{\substack{
h\in[M,\underline{BH}],\\
p(h)=g, }} 
e^{-2\pi i \int_M h^*x}
\end{equation} which has the required anomaly.

\subsection{Symmetry breaking as symmetry extension}
Let us now consider an extreme case of the construction in the last subsection.
Recall that one way to trivialize a class in $H^*(BG,U(1))$ is to consider the fibration \begin{equation}
G \to EG \to BG.
\end{equation} 
Since $EG$ is contractible, every class in $H^*(BG,U(1))$ trivializes when pull-backed via the projection.
This means that a boundary sigma model with the target space $G$ can couple to any bulk theory with $G$-symmetry.
In particular, the $G$-bundle trivializes on  the boundary.
This is the limiting case when $p=0$ in the discussion in the last subsection, in particular around \eqref{cf},
since for a finite group $G$ we have $K(G,0)=G$ and $K(G,1)=BG$.

Since the symmetry $G$ acts on the sigma model by a permutation, this corresponds to the symmetry breaking.
In general the dimension of the Hilbert space on $S^{d-1}$ is $|G|$.
We are interested in gapped boundaries where the symmetry is unbroken.
One necessary condition then is that the Hilbert space on $S^{d-1}$ is one-dimensional.

In the case of higher symmetries, we have a standard fibration\footnote{For a pointed topological space $X$ there is a path fibration $\Omega X\to LX \to X$, where $\Omega X$ is the loop space of $X$ and $LX$ is the path space of $X$ which is contractible. We also have $\Omega K(A,p+1) \cong K(A,p)$.} 
\begin{equation}
K(A,p) \to * \to K(A,p+1).
\end{equation}
Therefore, any class $H^*(K(A,p+1),U(1))$ characterizing the anomaly of a $p$-form $A$-symmetry is trivialized if we introduce a gauge field $\in C^{p}(-,A)$ for the  $(p{-}1)$-form $A$-symmetry on the boundary.
Again this corresponds to the spontaneous symmetry breaking of the $p$-form $A$-symmetry.
Note that this still keeps the Hilbert space on $S^{d-1}$ to be one dimensional.
So, in the case of the higher symmetry, we would like to keep the $A$ symmetry unbroken,
but this cannot be characterized by the dimension of the Hilbert space on $S^{d-1}$.
We note that the construction of the boundary theory describing the symmetry breaking of a higher symmetries was also discussed in a recent paper \cite{Hsin:2019fhf}.

\subsection{Cases beyond group cohomology}
\paragraph{Preliminaries:}
We next discuss how the symmetry extension method can be applied to SPT phases beyond group cohomology.
For definiteness we first consider the case of fermionic SPT phases specified  by $\omega\in \Hom(\Omega_{d+1}^\text{spin}(BG)^\text{tors},U(1))$, 
but the generalization to other cases should be straightforward and will be outlined at the end of this paper.
Suppose we have an extension \eqref{extension} such that $p^*(\omega) =1$, where $1$ here means the identity map sending any element to $1\in U(1)$.
How do we construct a $K$-gauge theory on the boundary, which produces for us a gapped boundary?

Consider a $(d+1)$-dimensional spin manifold $N_{d+1}$ with boundary $\partial N_{d+1} = M_{d}$, and its structure map $g : M_d \to BG$.
We would like to define a $K$-gauge theory on $M_{d}$ coupled with the $G$-background $g$.
This means that we would like  to sum over $h\in [M_d,BH] $ lifting $ g\in [M_d,BG]$, i.e. over $h$ such that $p(h)=g$,  so that 
we can define the partition function by\footnote{Here we assumed that $K$ is an abelian group for simplicity.}
\begin{equation}
	Z_\text{gauged}^K(M_d,g) \propto \sum_{p(h)=g} P(h).
    \label{eq:sumP}
\end{equation}
How do we define a phase $P(h)$ for each $h$?
Note that a state-sum definition of $\omega$ is not in general available.
Therefore we need to be slightly more abstract.

For this purpose, we use the Atiyah-Segal description of the invertible $G$-equivariant TQFT $Z^\omega_G$ associated to the anomaly $\omega$.  
For the details concerning  the construction of the invertible TQFT from the cobordism class $\omega\in\Hom(\Omega^{\spin}_{d+1}(BG),U(1))$, see \cite{Freed:2016rqq,Yonekura:2018ufj}.
We recall only the minimal information about it.
For a $d$-dimensional spin manifold $M_{d}$ equipped with a structure map $g\in [M_{d},BG]$, the TQFT assigns the Hilbert space $Z^\omega_G(M_{d},g)$. 
Because $Z^\omega_G$ is invertible, this Hilbert space is one-dimensional.
For a $(d+1)$-dimensional spin manifold $N_{d+1}$ with boundary $M_d\sqcup \overline{M}_d'$ and a map $\hat g\in [N_{d+1},BG]$,
the invertible TQFT assigns an isomorphism between Hilbert spaces
\begin{equation}
    Z^\omega_G(N_{d+1},\hat g):Z^\omega_G(M_{d},\hat g|_{M_{d}}) \to Z^\omega_G(M_{d}',\hat g|_{M_{d}'}),
\end{equation}
which is interpreted as the Euclidean time evolution along the manifold $N_{d+1}$ and the symmetry insertion $\hat g$.
We regard the empty set $\varnothing$ to be a $d$-dimensional (spin) manifold for any $d$.
A $(d{+}1)$-dimensional closed manifold equipped with a map $(N_{d+1},\hat{g})$ can be thought as a bordism between two empty sets. Then, the isomorphism $Z_G^\omega (N_{d+1},\hat{g}):Z^\omega_G(\varnothing)\to Z^\omega_G(\varnothing)$ provided by the invertible TQFT should be the multiplication by $\omega(N_{d+1},\hat{g})$.

\paragraph{An abstract construction:}
We first note that that a $d$-dimensional theory has an anomaly specified by $\omega$ is that 
the partition function of the boundary theory takes values in the one-dimensional vector space $Z^\omega_G(M_d,g)$ rather than in $\mathbb{C}$ with a canonically defined basis vector.
Therefore, in the partition function of the form \eqref{eq:sumP},
the phase $P(h)$ is better interpreted as a vector $\ket{P(h)}\in Z^\omega_G(M_d,g)$ whose norm is one,
and we need to provide a rule to find $\ket{P(h)}$.
The rule is provided by the assumption that $\omega$ trivializes when pulled back to $H$.
Indeed, our assumption is that $Z^{p^*(\omega)}_H$ is a trivial theory.
This means that there is a canonical basis vector in each of the 1-dimensional vector space:\begin{equation}
\ket{1}_{(M_d,h)}\in Z^{p^*(\omega)}_H(M_d,h),
\end{equation}
such that they are sent to themselves by the morphisms $Z^{p^*(\omega)}_H(N_{d+1},\hat h)$, etc.
That we obtained the $H$-symmetric theory by a pull-back provides an isomorphism \begin{equation}
\eta:Z^{p^*(\omega)}_H(M_d,h) \stackrel{\sim}{\longrightarrow} Z^\omega_G(M_d,p(h))=Z^\omega_G(M_d,g)
\end{equation}
and then we define \begin{equation}
\ket{P(h)}=\eta(\ket{1}_{(M_d,h)}) \in Z^\omega_G(M_d,g).
\end{equation}
This construction might sound  too abstract, so let us spell out the details.

\paragraph{A more concrete version:}
We first note that, because we assume $p^*(\omega) = 0$, $Z_G^\omega (L_{d+1},p(\hat{h}))$ is the identity map for any closed manifold $L_{d+1}$ and any map $\hat{h}$ to $BH$.
More generally, two bordisms $(L_{d+1},p(\hat{h}))$ and $(L_{d+1}',p(\hat{h}'))$ between manifolds $(M_{d},p(h))$ and $(M_{d}',p(h'))$ give the same map \begin{equation}
    Z_G^\omega(L_{d+1},p(\hat{h})) = Z_G^\omega(L_{d+1}',p(\hat{h}')):Z_G^\omega(M_{d},p(h))\to Z_G^\omega(M_{d}',p(h')),
\end{equation}
as long as all the involved structure maps can be lifted to $BH$. This can be shown by considering the union $L_{d+1}\cup L_{d+1}'$ and applying the statement about closed manifold.

Now, we construct the phase $P(h)$ in \eqref{eq:sumP} given the null-bordism $(N_{d+1},\hat{g})$ of the pair $(M_d,g)$ and a lift $h$ of $g$ with $p(h)=g$.
First, we arbitrarily fix vectors $\ket{0}_\varnothing\in Z_G^\omega(\varnothing)$ and $\ket{0}_{(M_d,g)}\in Z_G^\omega(M_d, g)$, and a lift $h_0$ of $g$ with $p(h_0)=g$. 
(If $g$ does not lift, the partition function \eqref{eq:sumP} is set to be zero.)
When $(M_d,h)$ and $(M_d,h_0)$ are bordant inside $BH$, 
we can set the phase $P$ as 
\begin{equation}
    P(h) = {}_{(M_d,g)}\!\bra{0}Z_G^\omega(L_{d+1}, p(\hat{h}))Z_G^\omega( N_{d+1}, \hat{g})\ket{0}_\varnothing,
    \label{eq:defP0}
\end{equation}
where $(L_{d+1}, \hat{h})$ is any bordism between $(M_d,h)$ and $(M_d,h_0)$.
This phase $P(h)$ has the required anomaly $\omega$ since the construction relies on an arbitrary choice $\ket{0}_{(M_d,g)}$.

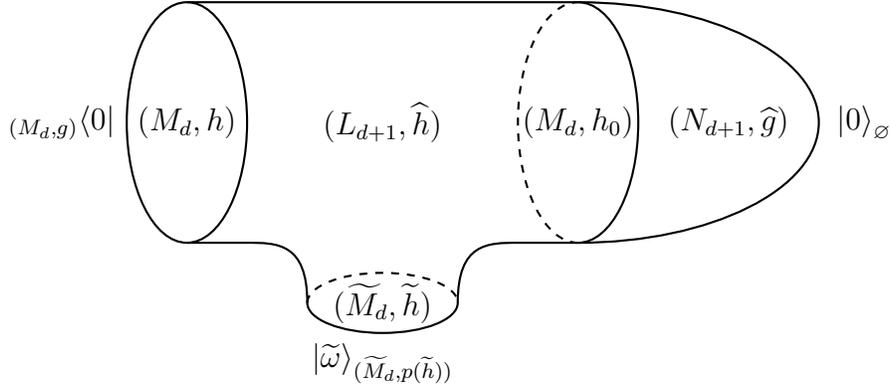
\begin{figure}
\center
\begin{tikzpicture}[scale = .8,thick]
\draw (0,0) coordinate (u) arc(90:-90:1 and 2) coordinate (d);
\draw[dashed] (d) arc(270:90:1 and 2);
\node (m) at ($(u)!.5!(d)$) {$(M_d,h_0)$};
\node (mr) at ($(m)+(2.5,0)$) {$(N_{d+1},\hat{g})$};
\node (mr2) at ($(mr)+(2.25,0)$) {$\ket{0}_{\varnothing}$};
\draw (u) arc(90:-90:4 and 2);
\draw (u) -- ++(-6.5,0) coordinate(ul4);
\draw (d) -- ++(-1,0) .. controls ++(-.5,0) and ++(0,1) .. ++(-1,-1) coordinate (dl) arc(0:-180: 1.25 and .5) coordinate (dl3) .. controls ++(0,1) and ++(.5,0) .. ++(-1,1) -- ++(-1,0) coordinate (dl4);
\draw[dashed] (dl3) arc(180:0:1.25 and .5);
\draw (ul4) arc(90:450:1 and 2);
\node (ml4) at ($(ul4)!.5!(dl4)$) {$(M_d,h)$};
\node[anchor=east] (ml5) at ($(ml4)+(-1,0)$) {${}_{(M_d,g)}\!\bra{0}$};
\node (mll) at ($(u)!.5!(ul4)+(0,-2)$) {$(L_{d+1},\hat{h})$};
\node (dl2) at ($(dl)!.5!(dl3)$) {$(\tilde{M}_d,\tilde{h})$};
\node (ddl3) at ($(dl2)+(0,-1)$) {$\ket{\tilde{\omega}}_{(\tilde{M}_d,p(\tilde{h}))}$};
\end{tikzpicture}
\caption{The geometric configuration defining $P(h)$ in \eqref{eq:defP}.}
\label{fig:Ph}
\end{figure}

For $h$ with which $(M_d,h)$ is not bordant to $(M_d,h_0)$ in $BH$, we need to introduce an additional state as depicted in Figure~\ref{fig:Ph}.
We choose a representative $(\tilde{M}_d^a,\tilde{h}^a)$ for each bordism class $a\in \Omega^{\spin}_d(BH)$,
and pick states $\ket{1}_{(\tilde{M}_d^a,p(\tilde{h}^a))} \in Z_G^\omega(\tilde{M}_d^a,p(\tilde{h}^a))$ satisfying the condition
\begin{equation}
    {}_{(\tilde{M}_d^{a+b},p(\tilde{h}^{a+b}))}\!\bra{1}Z_G^\omega(L_{d+1}^{a,b},p(\tilde{h}^{a,b}))\left(\ket{1}_{(\tilde{M}_d^a,p(\tilde{h}^a))}\otimes\ket{1}_{(\tilde{M}_d^b,p(\tilde{h}^b))}\right)=1,
    \label{eq:ket1}
\end{equation}
where $(\tilde{L}^{a,b}_{d+1},\tilde{h}^{a,b})$ is an arbitrary bordism between $(\tilde{M}_d^a,\tilde{h}^a)\sqcup (\tilde{M}_d^b,\tilde{h}^b)$ and $(\tilde{M}_d^{a+b},\tilde{h}^{a+b})$.

Such a choice of $\ket{1}_{(\tilde{M}_d^a,p(\tilde{h}^a))}$ is not unique, 
and another choice can be parametrized by $\tilde{\omega}\in\Hom(\Omega^\spin_d(BH),U(1))$ as
 \begin{equation}
     \ket{\tilde{\omega}}_{(\tilde{M}_d^a,p(\tilde{h}^a))}=\tilde{\omega}(\tilde{M}_d^a,\tilde{h}^a)\ket{1}_{(\tilde{M}_d^a,p(\tilde{h}^a))}.
 \end{equation}
Then, the phase $P(h)$ in general can be defined by 
\begin{equation}
    P(h) = {}_{(M_d,g)}\!\bra{0}Z_G^\omega(L_{d+1},p(\hat{h}))\left(\ket{\tilde{\omega}}_{(\tilde{M}_d,p(\tilde{h}))}\otimes Z_G^\omega( N_{d+1}, \hat{g})\ket{0}_\varnothing\right)
    \label{eq:defP}
\end{equation}
as illustrated in Figure~\ref{fig:Ph}, where $(\tilde{M_d},\tilde{h})$ is the chosen representative of the bordism class $[M_d,h]-[M_d,h_0]$ and $(L_{d+1},\hat{h})$ is an arbitrary bordism between $(M_d,h)$ and $(M_d,h_0)\sqcup (\tilde{M_d},\tilde{h})$. 
We have obtained multiple boundary theories in general, each of which (non-canonically) corresponds to an element $\tilde{\omega}$ of $\Hom(\Omega^\spin_d(BH),U(1))$;
they form a torsor over $\tilde{\omega}$ of $\Hom(\Omega^\spin_d(BH),U(1))$.
\footnote{Difference in the pure gravity part $\Hom(\Omega^\spin_d(\mathrm{pt}),U(1))$ of $\Hom(\Omega^\spin_d(BH),U(1))$ merely gives the difference in the gravitational counter term on the boundary.}

\paragraph{The case of cohomological SPT phases:}
The construction can be applied to the bosonic SPT phases by just ignoring the spin structure on the manifolds.
When the SPT corresponds to a cohomology element $\omega \in H^{d+1}(BG,U(1))$, the construction \eqref{eq:defP} coincides with the construction of \cite{Wang:2017loc} which was reviewed in Section~\ref{sec:extension}.
In this setup, when the $(M_d,h)$ and $(M_d,h_0)$ are bordant in $BH$, the formula \eqref{eq:defP0} computes
\begin{equation}
    P(h) \propto e^{-\int_{L_{d+1}} p(\hat{h})^*(\omega)+\int_{N_{d+1}}\hat{g}^*\omega},
    \label{eq:PWWW0}
\end{equation}
up to a overall phase independent of $h$. 
As $p^*\omega$ is trivial, we can take a cochain $x$ on $BH$ with $\delta x= p^*\omega$. Then the phase can be rewritten as 
\begin{equation}
    P(h) = e^{-\int_{M_{d}} h^*x + \int_{N_{d+1}}\hat{g}^*\omega},
    \label{eq:PWWW}
\end{equation}
where the factor $\int_{M_d}h_0^*x$ is absorbed into the overall coefficient.
The phase \eqref{eq:PWWW} is precisely what is reviewed in Section~\ref{sec:extension}.
Indeed, the formula \eqref{eq:PWWW} also holds  when $(M_d,h)$ and $(M_d,h_0)$ are not bordant.\footnote{The relation \eqref{eq:PWWW0} does not apply in this case, because the states $\ket{1}_{(\tilde{M}_d,\tilde{h})}$ are prepared so that they cancel the contribution $e^{\int_{\tilde{M}_d}\tilde{h}^*x}$ coming from $e^{\int_{L_{d+1}}p(\hat{h})^*\omega}$, which can be observed from \eqref{eq:ket1}.
}
The ambiguity parametrized by $\tilde{\omega}$ corresponds to the ambiguity of the choice of $x$.
\begin{figure}
\center
\begin{tikzpicture}[scale = .8,thick]
\draw (0,0) coordinate (u) arc(90:-90:1 and 2) coordinate (d);
\draw[dashed] (d) arc(270:90:1 and 2);
\node (m) at ($(u)!.5!(d)$) {$(M_d,0)$};
\node[anchor=west] (mr) at ($(m)+(1,0)$) {$\ket{0}_{(M_d,0)}$};
\draw (u) -- ++(-6.5,0) coordinate(ul4);
\draw (d) -- ++(-1,0) .. controls ++(-.5,0) and ++(0,1) .. ++(-1,-1) coordinate (dl) arc(0:-180: 1.25 and .5) coordinate (dl3) .. controls ++(0,1) and ++(.5,0) .. ++(-1,1) -- ++(-1,0) coordinate (dl4);
\draw[dashed] (dl3) arc(180:0:1.25 and .5);
\draw (ul4) arc(90:450:1.15 and 2);
\node (ml4) at ($(ul4)!.5!(dl4)$) {$(M_d,i(k))$};
\node[anchor=east] (ml5) at ($(ml4)+(-1,0)$) {${}_{(M_d,0)}\bra{0}$};
\node (mll) at ($(u)!.5!(ul4)+(0,-2)$) {$(L_{d+1},\hat{h})$};
\node (dl2) at ($(dl)!.5!(dl3)$) {$(\tilde{M}_d,\tilde{h})$};
\node (ddl3) at ($(dl2)+(0,-1)$) {$\ket{\tilde{\omega}}_{(\tilde{M}_d,p(\tilde{h}))}$};
\end{tikzpicture}
\caption{The geometric configuration defining $\omega_K$ in \eqref{eq:defomegaK}.}
\label{fig:omegaK}
\end{figure}
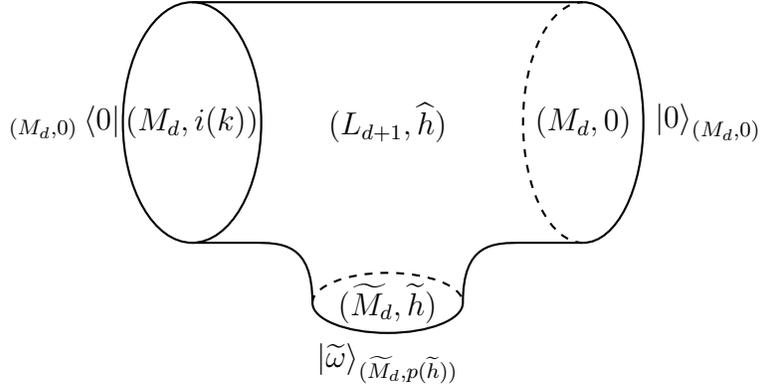

\paragraph{The action of the $K$-gauge theory:}
From the assumption $p^*(\omega)=1$, the $\omega$ should not contain the pure gravity anomaly. Therefore, the phase \eqref{eq:sumP} should define a spin $K$-gauge theory without gravitational anomaly. When the background $g$ is turned off, the action of such a gauge theory is supposed to be given by an element $\omega_K\in \Hom(\Omega^{\spin}_d(BK)^\mathrm{tors},U(1))$. However, the definition \eqref{eq:defP} works only when $M_d$ is null-bordant as a spin manifold without an additional structure map.
For general $M_d$, we can instead define $\omega_K$ as
\begin{equation}
    \omega_K(k\in [M_d,BK]) =  {}_{(M_d,0)}\!\bra{0}Z_G^\omega(L_{d+1},p(\hat{h}))\left(\ket{\tilde{\omega}}_{(\tilde{M}_d,p(\tilde{h}))}\otimes \ket{0}_{(M_d,0)}\right),
    \label{eq:defomegaK}
\end{equation}
where $i:K\to H$ is the injection with $p\circ i =0$, $(\tilde{M},\tilde{h})$ is the representative of the bordism class $[M_d,i(k)]-[M_d,0]$, and $(L_{d+1},\hat{h})$ is an arbitrary bordism between $(M_d,i(k))$ and $(M_d,0)\sqcup(\tilde{M}_d,\tilde{h})$.
The geometry of \eqref{eq:defomegaK} is illustlated in Figure~\ref{fig:omegaK}.
One can check that this action depends only on the spin-bordism class of $(M_d,k)$, and hence $\omega_K\in \Hom(\Omega_d^\spin(BK),U(1))$.
This action $\omega_K$ depends on $\tilde{\omega}\in\Hom(\Omega_d^\spin(BH),U(1))$, and in fact the set of possible $\omega_K$ obtained in this way is an  $i^*\Hom(\Omega_d^\spin(BH),U(1))$-torsor.
Therefore, we have obtained the secondary cohomology operation
\begin{equation}
    \mathrm{Ker}(p^*_{d+1}) \to \mathrm{Coker}(i^*_d)
\end{equation}
of the Pontryagin dual of the spin-bordism homology, where $p^*_{d+1}:\Hom(\Omega_{d+1}^\spin(BG),U(1)) \to \Hom(\Omega_{d+1}^\spin(BH),U(1))$ and $i^*_{d}:\Hom(\Omega_{d}^\spin(BH),U(1)) \to \Hom(\Omega_{d}^\spin(BK),U(1))$ are the pull-backed associated to $p:BH\to BG$ and $i:BK\to BH$.

\paragraph{Further generalizations:}
The construction above can be generalized further in a few ways.
First, the manifold structure does not necessarily have to be the spin structure. 
It could be  orientation, pin${}^\pm$, or nothing at all. In these cases we  use the corresponding bordism groups, namely oriented, pin${}^\pm$, or unoriented bordism group, respectively. 
A more exotic structure could be considered if we wanted.
Second, the groups $K$, $H$, and $G$ do not necessarily have to be ordinary groups, but can be higher-groups, because the construction works for a general fibration $F\to E \to B$ instead of $BK \to BH \to BG$ with ordinary groups $K$, $H$ and $G$.\footnote{Although $H$ and $G$ can also be continuous groups, $K$ needs to be a finite group, or $|\pi_*(BK)|$ needs to be finite when $K$ is a higher group, for the sum \eqref{eq:sumP} to make sense. When $K$ is not an ordinary group, the factor $\frac{1}{|K|^{|\pi_0(M_d)|}}$ should be modified. For the case with $BK=K(A,p)$, the factor is replaced by $\prod_{i=0}^{(p-1)/2} \frac{|H^{2i-1}(M_d,A)|}{|H^{2i}(M_d,A)|}$ when $p$ is odd, where $|H^{-1}(M_d,A)|$ is understood to be $1$, and $\prod_{i=0}^{p/2-1} \frac{|H^{2i}(M_d,A)|}{|H^{2i+1}(M_d,A)|}$ when $p$ is even. These factors represents the residual gauge redundancies, the gauge redundancies of the gauge redundancies, and so on. For the general $K$ case, see \cite{Freed:2018cec} and references therein.}

Lastly, we can generalize the construction by allowing for $H$ and $G$ to involve the spacetime symmetry.
In this case, the anomaly $\omega$ can contain a pure gravity part. For example, we can take the pure gravitational anomaly $\omega  \in \Hom(\Omega^{SO}_5(\mathrm{pt}),U(1))$
which corresponds to the Stiefel-Whitney polynomial $w_2w_3$.
This anomaly can be trivialized by the extension of the spacetime symmetry group
\begin{equation}
    1 \to \mathbb{Z}_2 \to Spin(5) \to SO(5) \to 1.
\end{equation}
In this case, $\Omega^\spin_{d+1}(BG)$ and $\Omega^\spin_{d+1}(BH)$ in the above construction is replaced by $\Omega^{SO}_{d+1}(\mathrm{pt})$ and $\Omega^{\spin}_{d+1}(\mathrm{pt})$.
Then the boundary theory is given by summing over the spin structures on $M_d$ with the phase \eqref{eq:sumP}.\footnote{When $\omega$ contains a pure gravity part, there is no way to ``turn off'' the gravity background.
So we cannot extract a purely $d$-dimensional $K$-gauge theory from the construction, and hence the paragraph containing \eqref{eq:defomegaK} does not generalize to this case.}
As a cochain field theory, it can be written as \begin{equation}
Z \propto \sum_{\delta a=w_2} \exp(i\int_M a\cup w_3)  
= \sum_{\substack{a\in C^1(M,\bZ_2),\\ b\in C^{d-2}(M,\bZ_2)}}
 \exp(\pi i\int_M \left(a \cup \delta b + a\cup w_3 + w_2\cup b\right)),
\end{equation}
and was discussed in \cite{Thorngren:2014pza}.
In general, when $H$ and $G$ involve the spacetime symmetry, the bordism group of $BH$ and $BG$ needs to be replaced by the corresponding Madsen-Tilman spectra. See \cite{Freed:2016rqq} for detail.

\paragraph{Summary:}
Summarizing, given a symmetry extension trivializing the anomaly, the construction \eqref{eq:defP} provides an abstruct construction of the topological boundary theory of the corresponding SPT. 
In the rest of the paper, we are going to give more concrete constructions for certain cases. 
In Sec.~\ref{sec:unoriented}, we will see that the cochain integrals are sufficient for our purposes for time-reversal-invariant bosonic SPT phases.
In Sec.~\ref{sec:GuWen}, we will discuss the use of the Gu-Wen Grassmann integral in the case of the Gu-Wen SPT phases later.

\section{For time-reversal invariant bosonic SPT phases}
\label{sec:unoriented}

In this section we discuss time-reversal-invariant bosonic SPT phases protected by a finite symmetry group $G$, where we assume that the $G$ action and the time-reversal action are independent.
We can call them as unoriented bosonic invertible phases, and they are described by $\Hom(\Omega^\text{unoriented}_{d+1}(X),U(1))$ where $X=BG$.
Luckily, an explicit and complete description of this group was already given in the algebraic topology literature in the 1960s \cite{BrownPeterson}\footnote{%
The 2nd edition of the textbook \cite{Conner} contains a very readable account in its Chapter I.18.}
This allows us to construct gapped boundaries for all of them.

We first recall the  homomorphism \begin{equation}
H^*(BO,\bZ_2)\otimes H^*(X,\bZ_2) \to \Hom(\Omega^\text{unoriented}_{d+1}(X),U(1)).\label{unopoly}
\end{equation}
This is obtained by integrating an element on the left hand side, 
i.e.~a polynomial of the universal Stiefel-Whitney classes $w_i$ and the cohomology classes $\alpha_i$ of $X$, 
on the $(d+1)$-dimensional manifold $M$ equipped with a map $f$ to $X$, 
by using the Stiefel-Whitney classes $w_i(TM)$ of the tangent bundle  and the pullbacks $f^*(\alpha_i)$.
The theorem \cite{BrownPeterson} asserts that this homomorphism is surjective;
the theorem also explicitly describes the kernel.

Let us now show that we can  construct a gapped boundary theory for an unoriented invertible phase by the symmetry extension.
We already know that the symmetry  extension allows us to kill any cohomology class in $H^{q\ge 2}(BG,\bZ_2)$,
since we assumed that $G$ is finite.
Therefore we can assume without loss of generality that the bulk invertible phase is specified by \begin{equation}
P_{d+1}(w_i) + \alpha_1 Q_{d}(w_i), \label{PQ}
\end{equation} where $\alpha_1\in H^1(BG,\bZ_2)$ and $P,Q\in H^*(BO,\bZ_2)$ with the degrees specified in the subscripts.

We note that introducing a $p$-form $\bZ_2$ gauge field $a$ on the boundary with $\delta a=w_{p+2}$ corresponds to an extension of the structure
\begin{equation}
K(\bZ_2, p+1)\to \underline{BH}\to BO,
\end{equation}
and trivializes the anomaly involving $w_{p+2}$. 
(We note that we prefer to take $p\le d-3$. If $p=d-2$, the Hilbert space on $S^{d-1}$ can be two-dimensional, which we do not want.)
Therefore the question is whether we can trivialize the entire anomaly \eqref{PQ} by repeating this process.

This can be done recursively, as follows.
We use a mathematical result \cite{PengelleyWilliams} which says that $H^*(BO,\bZ_2)$ as an algebra over the Steenrod algebra is generated by $w_1$, $w_2$, $w_4$, \ldots, $w_{2^r}$,\ldots.
This in particular means that if $w_{2^r}=0$ for $r\le R$, we have $w_{i}=0$ for $i<2^{R+1}$, since these $w_i$ can be generated from $w_{2^r}$ with $r\le R$ using the Steenrod squares, additions and multiplications.
We also use the fact that the Wu class has the form \begin{equation}
\nu_{2^r}=w_{2^r} + \text{decomposables}
\end{equation}
and that the Wu class $\nu_k$ vanishes on a $(d+1)$-dimensional space if $2k\ge d+1$; for Wu classes, see e.g.~\cite{MS} or \cite{ManifoldAtlasWu}.

First, we introduce two $1$-form $\bZ_2$ gauge fields (for $0$-form $\bZ_2$ gauge symmetries) $a,b$ trivializing $(w_1)^2$ and $w_2$.
This kills all polynomials of Stiefel-Whitney classes up to and including $w_3$, already at the level of $H^*(BO,\bZ_2)$.
Since the Wu class $\nu_4$ vanishes if $d+1<8$, 
$w_4$ also vanishes, and therefore
every Stiefel-Whitney polynomial (except $w_1$ itself) vanishes and the anomaly \eqref{PQ} is trivialized if $d+1<8$.

Next, when $d+1\ge 8$,
we  introduce a $3$-form $\bZ_2$ gauge field $a_3$  trivializing $w_4$.
This kills all polynomials of Stiefel-Whitney classes up to and including $w_7$, already at the level of $H^*(BO,\bZ_2)$.
Since the Wu class $\nu_8$ vanishes if $d+1<16$, 
$w_8$ also vanishes, and therefore
every Stiefel-Whitney polynomial (except $w_1$ itself) vanishes and the anomaly \eqref{PQ} is trivialized if $d+1<16$.

In general, when $d+1\ge 2^{r+1}$,
we  introduce a $(2^r-1)$-form $\bZ_2$ gauge field $a_{2^r-1}$  trivializing $w_{2^r}$.
This kills all polynomials of Stiefel-Whitney classes up to and including $w_{2^{r+1}-1}$, already at the level of $H^*(BO,\bZ_2)$.
Since the Wu class $\nu_{2^{r+1}}$ vanishes if $d+1<2^{r+2}$, 
$w_{2^{r+1}}$ also vanishes, and therefore
every Stiefel-Whitney polynomial (except $w_1$ itself) vanishes and the anomaly \eqref{PQ} is trivialized if $d+1<2^{r+2}$.

\section{For Gu-Wen spin SPT phases}
\label{sec:GuWen}
The Gu-Wen phases are a subset of fermionic SPT phases which admit a particularly explicit description, first studied in \cite{Gu:2012ib} and further explored in \cite{Gaiotto:2015zta}.
The aim of this section is to construct gapped boundaries for Gu-Wen phases by the symmetry extension method.
As we will see, the applicability of this method requires that we can trivialize the cohomology classes specifying the Gu-Wen phase by some extension.
This condition is automatically satisfied for any finite group $G$,
and therefore our methods provides a gapped boundary for an arbitrary Gu-Wen phase for any finite group $G$.

\subsection{Strategy}
The Gu-Wen spin invertible theories form a subgroup of $\Hom(\Omega^\text{spin}_{d+1}(BG),U(1))$ and is specified by a pair $(n_d,y_{d+1})\in Z^{d}(BG,\bZ_2)\times C^{d+1}(BG,U(1))$ satisfying $\Sq^2 n_d = \delta y_{d+1}$, where $\Sq^2 n := n \cup_{d-2} n$.
For a given $g:N\to BG$ where $N$ is a spin $(d+1)$-manifold, the action of the invertible theory is given by \cite{Gu:2012ib,Gaiotto:2015zta}\footnote{%
For a more mathematical treatment, see papers by Brumfiel and Morgan \cite{BFquadratic}.
} \begin{equation}
\sigma(g^*n_d)  \exp(\pi i \int_N(\eta\cup g^*n_d + g^*y_d))
\end{equation}
where $\sigma(g^*n_d)=\pm1$ is the Grassmann integral of Gu-Wen \cite{Gu:2012ib} as formulated by Gaiotto and Kapustin \cite{Gaiotto:2015zta},
and $\delta\eta=w_2$ specifies the chosen spin structure.\footnote{
In \cite{Gaiotto:2015zta} Gaiotto and Kapustin proposed and used an explicit cocycle representative of $w_2$.
We note that the explicit cocycle representatives  for $w_n$ were in fact originally conjectured by Stiefel and Whitney; this was later proved in the 70s, see e.g. p.143 of \cite{MS}, or \cite{HalperinToledo,BlantonMcCrory} and references therein.
}

In this subsection, we  write down the explicit $d$ dimensional action on the boundary of $(d+1)$ dimensional Gu-Wen spin $G$-SPT phase specified by the Gu-Wen data $(n_d,y_{d+1})$.
To construct the gapped boundary, we prepare a symmetry extension by a symmetry $\tilde{K}$
\begin{equation}
0\to \tilde{K}\to \tilde{H} \stackrel{\tilde{p}}{\to} G \to 0 \label{eq:1st-ext}
\end{equation} 
such that $n_d$ trivializes as an element of $H^d(B\tilde{H}, \bZ_2)$; 
$[\tilde{p}^*n_d]=0\in H^d(B\tilde{H}, \bZ_2)$. 

We now take $\tilde{m}_{d-1}\in C^{d-1}(B\tilde{H}, \bZ_2)$ such that $\tilde{p}^* n_d=\delta \tilde{m}_{d-1}$.
We see that $z_{d+1}=\tilde{p}^*y_{d+1}-\Sq^2\tilde{m}_{d-1}$ is a cocycle, where $\Sq^2 \tilde{m}_{d-1}=\tilde{m}_{d-1}\cup_{d-3} \tilde{m}_{d-1} + \delta \tilde{m}_{d-1} \cup_{d-2} \tilde{m}_{d-1}$.
Therefore the bulk Gu-Wen data are pulled back to $(\delta \tilde{m}_{d-1}, \Sq^2 \tilde{m}_{d-1}+ z_{d+1})$.
We now assume that there is a further extension of the symmetry 
\begin{equation}
0\to K \to  H  \stackrel{p}{\to}  \tilde{H} \to 0 \label{eq:2nd-ext}
\end{equation}
such that $p^*z_{d+1}=\delta x_d$ for some $x_d\in C^{d}(BH,\U(1))$.
We set $m_{d-1}=p^*\tilde{m}_{d-1}$.

When $G$ is finite, the necessary extensions \eqref{eq:1st-ext} and \eqref{eq:2nd-ext} can be prepared by generalizing the argument of \cite{Tachikawa:2017gyf}.
In the general discussion below, we simply need such an extension, possibly with a higher-form symmetry, so that the cohomology classes involved trivialize.\footnote{The degrees of these higher form symmetries $K$ and $\tilde{K}$ need to be less than $d-2$ for the resulting gauge theory to be meaningful as a TQFT. When either $K$ or $\tilde{K}$ is of degree $d-2$, the TQFT has degenerate vacua on $S^{d-1}$. When $K$ and $\tilde{K}$ have different degrees, the boundary TQFT becomes in general a gauge theory with a higher-group, which is more general than the higher-form symmetry, and the gauge redundancy factor of such a theory should be taken care of carefully \cite{Freed:2018cec}.}

We now expect that the action of the $K$-gauge theory on the boundary is given by \begin{equation}
\sum_{p(h)=g} \sigma(h^*m_{d-1}) \exp(\pi i \int_{M} (\eta\cup h^*m_{d-1} + h^*x_d)),
\label{gwb}
\end{equation}
but to make sense of this expression  we have to extend the definition of the Gu-Wen Grassmann integral $\sigma(\alpha_{d-1})$ to the case when $\alpha_{d-1}\in C^{d-1}(M, \bZ_2)$ is not necessarily closed. 
We will see that such extended Gu-Wen integral nicely couples to the bulk in a gauge invariant fashion.
For this purpose, let us start by recalling the construction of the Gu-Wen Grassmann integral $\sigma(M, \alpha)$ for closed $\alpha$.

\subsection{Review of the Gu-Wen Grassmann integral}
We first endow $M$ with a triangulation. In addition, we take the
barycentric subdivision for the triangulation of $M$. Namely, each $d$-simplex in the initial triangulation of $M$ is subdivided into $(d+1)!$ simplices, whose vertices are barycenters of the subsets of vertices in the $d$-simplex. We further assign a local ordering to vertices of the barycentric subdivision, such that a vertex on the barycenter of $i$ vertices is labeled by $i$,
as was done in \cite{Gaiotto:2015zta}.
Each simplex can then be either a $+$ simplex or a $-$ simplex, depending on whether the ordering agrees with the orientation or not.
We assign a pair of Grassmann variables $\theta_e, \overline{\theta}_e$ on each $(d-1)$-simplex $e$ of $M$ such that $\alpha(e)=1$ for a given $\alpha \in Z^{d-1}(M,\bZ_2)$. 
We say that $\theta_e$ is contained in one of  $d$-simplices neighboring $e$, and $\overline{\theta}_e$ is contained in the other $d$-simplex;
we will specify the detail soon.
Then, $\sigma(M, \alpha)$ is defined as
\begin{equation}
    \sigma(M, \alpha)=\int\prod_{e|\alpha(e)=1}\! d\theta_e d\overline{\theta}_e\  \prod_t u(t),
    \label{sigmadef}
\end{equation}
where $t$ denotes a $d$-simplex, and $u(t)$ is the product of Grassmann variables contained in $t$.
For instance, for $d=2$, $u(t)$ on $t=(012)$ is the product of
$\vartheta_{12}^{\alpha(12)}, \vartheta_{01}^{\alpha(01)}, \vartheta_{02}^{\alpha(02)}$. 
Here, $\vartheta$ denotes $\theta$ or $\overline{\theta}$ depending on the choice of the assigning rule, which will be discussed later. The order of Grassmann variables in $u(t)$ will also be defined shortly.
We note that $u(t)$ is ensured to be Grassmann-even when $\alpha$ is closed. 

\begin{figure}[htb]
\centering
\includegraphics{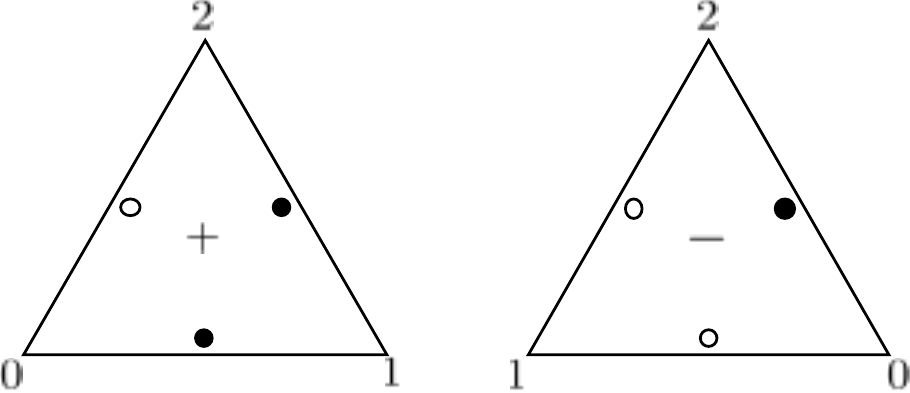}
\caption{Assignment of Grassmann variables on 1-simplices in the case of $d=2$. $\theta$ (resp.~$\overline{\theta}$) is represented as a black (resp.~white) dot.}
\label{fig:Grassmann}
\end{figure}

Due to the fermionic sign of Grassmann variables, $\sigma(\alpha)$ becomes a quadratic function, whose quadratic property depends on the order of Grassmann variables in $u(t)$. We will adopt the order used in Gaiotto-Kapustin \cite{Gaiotto:2015zta}, which is defined as follows. 
\begin{itemize}
\item
For $t=(01\dots d)$, we label a $(d-1)$-simplex $(01\dots\hat{i}\dots d)$ (i.e.~a $(d-1)$-simplex given by omitting a vertex $i$) simply as $i$. 
\item Then, the order of $\vartheta_i = \vartheta_{01\cdots \hat{i} \cdots d}$ for $+$ $d$-simplex $t$ is defined by first assigning even $(d-1)$-simplices in ascending order, then odd simplices in ascending order again:
\begin{equation}
    0\to 2\to 4\to\dots \to 1\to 3\to 5\to\dots
\end{equation}
\item For $-$ $d$-simplices, the order is defined in opposite way:
\begin{equation}
    \dots\to 5\to 3\to 1 \to \dots \to 4\to 2\to 0.
\end{equation}
\end{itemize}
For example, for $d=2$, $u(012)=\vartheta_{12}^{\alpha(12)}\vartheta_{01}^{\alpha(01)}\vartheta_{02}^{\alpha(02)}$ when $(012)$ is a $+$ triangle, 
and $u(012)=\vartheta_{02}^{\alpha(02)}\vartheta_{01}^{\alpha(01)}\vartheta_{12}^{\alpha(12)}$ for a $-$ triangle. 
Then, we choose the assignment of $\theta$ and $\overline{\theta}$ on each $e$ such that  $u(t)$ includes $\overline{\theta}_e$ when $e$ is labeled by an odd (resp.~even) number  if $t$ is a $+$ (resp.~$-$) simplex, see Fig.~\ref{fig:Grassmann}.

Based on the above definition of $u(t)$, the quadratic property of $u(t)$ is given by
\begin{equation}
    \sigma(\alpha)\sigma(\alpha')=\sigma(\alpha+\alpha')(-1)^{\int\alpha\cup_{d-2}\alpha'},
    \label{eq:ClosedQuad}
\end{equation}
for closed $\alpha, \alpha'$. To see this, we just have to bring the product of two Grassmann integrals
\begin{equation}
    \sigma(\alpha)\sigma(\alpha')=\int\prod_{e|\alpha(e)=1}d\theta_e d\overline{\theta}_e\prod_{e|\alpha'(e)=1}d\theta_e d\overline{\theta}_e \prod_t u(t)[\alpha]\prod_t u(t)[\alpha']
\end{equation}
into the form of $\sigma(\alpha+\alpha')$ by permuting Grassmann variables, and count the net fermionic sign.
First of all, each path integral measure on $e$ picks up a
sign $(-1)^{\alpha(e)\alpha'(e)}$ by permuting $d\overline{\theta}_e^{\alpha(e)}$ and $d\theta_e^{\alpha'(e)}$.
For integrands, $u(t)$ on different $d$-simplices commute with each other for closed $\alpha$, so nontrivial signs occur only by reordering $u(t)[\alpha]u(t)[\alpha']$ to $u(t)[\alpha+\alpha']$ on a single $d$-simplex. The sign on $t$ is explicitly written as
\begin{equation}
    (-1)^{\sum_{e,e'\in t}^{e>e'}\alpha(e)\alpha'(e')},
\end{equation}
where the order $e>e'$ is determined by $u(t)$. Hence, the net fermionic sign is given by
\begin{equation}
    \sigma(\alpha)\sigma(\alpha')=\sigma(\alpha+\alpha')\prod_t(-1)^{\epsilon[t, \alpha, \alpha']},
    \label{eq:orientedquad}
\end{equation}
with
\begin{equation}
   \epsilon[t, \alpha, \alpha']=\sum_{e,e'\in t, e>e'}\alpha(e)\alpha'(e')+\sum_{e\in t, e>0}\alpha(e)\alpha'(e),
   \label{eq:redis}
\end{equation}
where $e>0$ if $u[t]$ includes a $\overline{\theta}_e$ variable. Then, the sign $\epsilon[t, \alpha, \alpha']$ has a neat expression in terms of the higher cup product.
For later convenience, we compute $\epsilon[t, \alpha, \alpha']$ including the case that $\alpha, \alpha'$ are not closed. 

At a $+$ simplex, after some efforts we can rewrite $\epsilon[t, \alpha, \alpha']$ as
\begin{equation}
\begin{split}
    \epsilon[t, \alpha, \alpha']&=\sum_{i}\alpha_{2i+1}\cdot \delta\alpha'(t)+\sum_{i<j}\alpha_{2i+1}\alpha'_{2j+1}+\sum_{i>j}\alpha_{2i}\alpha'_{2j}\\
    &=\alpha\cup_{d-2}\alpha'+\alpha\cup_{d-1}\delta\alpha'.
    \end{split}
    \label{eq:pepsilon}
\end{equation}

At a $-$ simplex, similarly we have
\begin{equation}
\begin{split}
    \epsilon[t, \alpha, \alpha']&=\sum_{i}\alpha_{2i}\cdot \delta\alpha'(t)+\sum_{i<j}\alpha_{2i+1}\alpha'_{2j+1}+\sum_{i>j}\alpha_{2i}\alpha'_{2j}\\
    &=\delta\alpha(t)\delta\alpha'(t)+\alpha\cup_{d-2}\alpha'+\alpha\cup_{d-1}\delta\alpha'.
    \end{split}
    \label{eq:mepsilon}
\end{equation}
We can see the quadratic property~\eqref{eq:ClosedQuad} when $\alpha, \alpha'$ are closed. 

The change of $\sigma(\alpha)$ under the gauge transformation
$\alpha\to \alpha+\delta \gamma$  or under the change of the triangulation is controlled by the formula \begin{equation}
\sigma(\tilde M,\tilde \alpha) 
= (-1)^{\int_K (\Sq^2 (\alpha) + w_2 \cup \alpha)} \sigma(M,\alpha), 
\end{equation}
where $\tilde M$ is the same manifold $M$ with a different triangulation,
$\tilde\alpha$ is a cocycle such that $[\alpha]=[\tilde\alpha] $ in cohomology,
and $K=M\times [0,1]$ such that the two boundaries are given by $M$ and $\tilde M$,
and finally $\alpha$ is extended to $K$ so that it restricts to $\alpha$ and $\tilde \alpha$ on the boundaries.
The derivation was given in \cite{Gaiotto:2015zta}.

We note that due to the Wu relation~\cite{ManifoldAtlasWu}, we have \begin{equation}
(-1)^{\int_K (\Sq^2(\alpha) + w_2\cup\alpha)} = +1,
\end{equation}  when $K$ is an oriented closed manifold and $\alpha$ is a cocycle.
This means that $\int_K (\Sq^2 (\alpha) + w_2\cup\alpha)$ represents a trivial phase in $d+1$ dimensions,
and therefore there should be a trivial boundary in $d$ dimensions.
We can think of the Gu-Wen Grassmann integral $\sigma(M,\alpha)$ as providing an explicit formula for such a trivial boundary.

\subsection{Bulk-boundary Gu-Wen Grassmann integral}
When we naively use the above definition \eqref{sigmadef} when  $\alpha$ is not closed: $\delta \alpha=\beta$, 
the resulting expression is problematic since $u(t)$ can become Grassmann-odd. 
We can avoid this conundrum by coupling it with the Gu-Wen integral $\sigma(N,\beta)$ in $(d+1)$ dimensional bulk $N$ such that $\partial N=M$, making all components in the path integral Grassmann-even. 

Now let us write down the boundary Gu-Wen integral coupled with bulk;
we denote the entire integral by $\sigma(\alpha;\beta)$. 
We assign Grassmann variables $\theta_e, \overline{\theta}_e$ on each $(d-1)$-simplex $e$ of $M$, and $\theta_f, \overline{\theta}_f$ on each $d$-simplex $f$ of $N\setminus M$.
We define the Gu-Wen integral as
\begin{equation}
    \sigma(\alpha;\beta)=\int\prod_{f|\beta(f)=1}d\theta_f d\overline{\theta}_f \int\prod_{e|\alpha(e)=1}d\theta_e d\overline{\theta}_e \prod_t u(t),
    \label{eq:GWboundary}
\end{equation}
where $u(t)$ is a monomial of Grassmann variables defined on a $(d+1)$-simplex of $N$. 
$u(t)[\beta]$ is defined in the same fashion as the case without boundary if $t$ is away from the boundary, but modified when $t$ shares a $d$-simplex with the boundary. 
For simplicity, we assign an ordering on vertices of such $t=(01\dots d+1)$, so that the $d$-simplex shared with $M$ becomes $f_0=(12\dots d+1)$; the vertex $0$ is contained in $N\setminus M$. 
For instance,  we can take a barycentric triangulation on $N$, and assign $0$ to vertices associated with $(d+1)$-simplices.  
Then, $u(t)$ with $t$ neighboring with $M$ is defined by replacing the position of $\vartheta_{f_0}$ in $u(t)[\beta]$ with the boundary action on $f_0$, $u(f_0)[\alpha]=\prod_{e\in f_0}\vartheta_e^{\alpha(e)}$. 
We then have: On a $+$ simplex,
\begin{equation}
    u(t)=u(f_0)[\alpha]\cdot\prod_{f\in\partial t, f\neq f_0}\vartheta_f^{\beta(f)}.
\end{equation}
On a $-$ simplex,
\begin{equation}
    u(t)=\prod_{f\in\partial t, f\neq f_0}\vartheta_f^{\beta(f)}\cdot u(f_0)[\alpha].
\end{equation}
One can check that $u(t)$ defined above becomes Grassmann-even.
For later convenience, we will also define the variant $\overline{\sigma}(\alpha;\beta)$ of the Gu-Wen integral $\sigma(\alpha;\beta)$ defined above, by changing the role of $\theta$ and $\overline{\theta}$ in $u(f_0)$.  Namely, we use $u(t)=\overline{u}(f_0)\cdot\prod_{f\in\partial t, f\neq f_0}\vartheta_f^{\beta(f)}$ in~\eqref{eq:GWboundary}, where $\overline{u}(f_0)$ denotes a monomial given by replacing $\theta\leftrightarrow\overline{\theta}$ in $u(f_0)$. $\sigma(\alpha;\beta)$ and $\overline{\sigma}(\alpha;\beta)$ only differs by linear and gauge invariant counterterm on $M$,
\begin{equation}
(-1)^{\sum_{e\in M} \alpha(e)}=(-1)^{\sum_{f_{+}\in M}\beta(f_{+})},
\end{equation}
where $f_{+}$ denotes $+$ simplices in $M$.

We now show that the modified Gu-Wen integrals~\eqref{eq:GWboundary} $\sigma, \overline{\sigma}$ both satisfy the quadratic property
\begin{equation}
    \sigma(\alpha+\alpha'; \beta+\beta')=\sigma(\alpha;\beta)\sigma(\alpha';\beta')(-1)^{\int_M \alpha\cup_{d-2}\alpha'+\alpha\cup_{d-1}\delta\alpha'+\int_N \beta\cup_{d-1}\beta'}.
    \label{eq:Quadboundary}
\end{equation}
Basically, the quadratic property is derived in the same manner as the case without boundary. The net fermionic sign is expressed in terms of
\begin{equation}
    \sigma(\alpha;\beta)\sigma(\alpha';\beta')=\sigma(\alpha+\alpha'; \beta+\beta')\prod_f(-1)^{\epsilon[f,\alpha, \alpha']}\prod_t(-1)^{\tilde{\epsilon}[t,\beta, \beta']}.
\end{equation}
Here, $\epsilon[f,\alpha,\alpha']$ is the same as~\eqref{eq:pepsilon},~\eqref{eq:mepsilon}, which counts the sign on the boundary; $(-1)^{\alpha(e)\alpha'(e)}$ by permuting the measure $d\overline{\theta}_e^{\alpha(e)}$, $d\theta_e^{\alpha'(e)}$ on $(d-1)$-simplices in $M$, and the sign that occurs by reordering $u(f_0)[\alpha]u(f_0)[\alpha']$ to $u(f_0)[\alpha+\alpha']$ on a $d$-simplex $f_0$ in $M$. 

$\tilde{\epsilon}[t,\beta,\beta']$ counts the sign on the bulk, which is identical to $\epsilon[t,\beta,\beta']$ away from the boundary,
that is $\tilde{\epsilon}[t,\beta,\beta']=\beta\cup_{d-1}\beta'$. 
However, when $t$ shares a $(d-1)$-simplex $f_0$ with $M$, the sign is modified at $-$ simplices due to the absence of $(-1)^{\beta(f_0)\beta'(f_0)}$ sign from the measure, since we do not have a Grassmann variable $\vartheta_{f_0}$ attached to $f_0$. 
Hence, on a $+$ simplex we have \begin{equation}
   \tilde{\epsilon}[t,\beta,\beta']=\beta\cup_{d-1}\beta'.
\end{equation}
However, on a $-$ simplex we instead have
\begin{equation}
   \tilde{\epsilon}[t,\beta,\beta']=\beta\cup_{d-1}\beta'-\beta(f_0)\beta'(f_0).
   \label{eq:mepsilontilde}
\end{equation}
Now, we see that on the boundary such that $f_0\in\partial t$,
\begin{equation}
    \epsilon[f_0,\alpha,\alpha']\tilde{\epsilon}[t,\beta,\beta']= \alpha\cup_{d-2}\alpha'+\alpha\cup_{d-1}\delta\alpha'+ \beta\cup_{d-1}\beta',
\end{equation}
on both $+$ and $-$ simplices. 
Here, we are choosing the orientation of simplices such that the orientation of a $d$-dimplex $f$ agrees with $t$ such that $f\in\partial t$, when $f$ is labeled by an even integer, and disagrees when $f$ is labeled by an odd integer. 
Then, we have the identical orientation on $f_0$ and $t$, hence in $-$ simplices the $\beta(f_0)\beta'(f_0)$ term in~\eqref{eq:mepsilontilde} cancels with the $\delta\alpha(f_0)\delta\alpha'(f_0)$ in~\eqref{eq:mepsilon}.
 Therefore, now we see that the overall fermionic sign is summarized as~\eqref{eq:Quadboundary}.

\subsection{Effect of the change of the triangulation}
\label{sec:retriangle}
\begin{figure}[tb]
\centering
\includegraphics{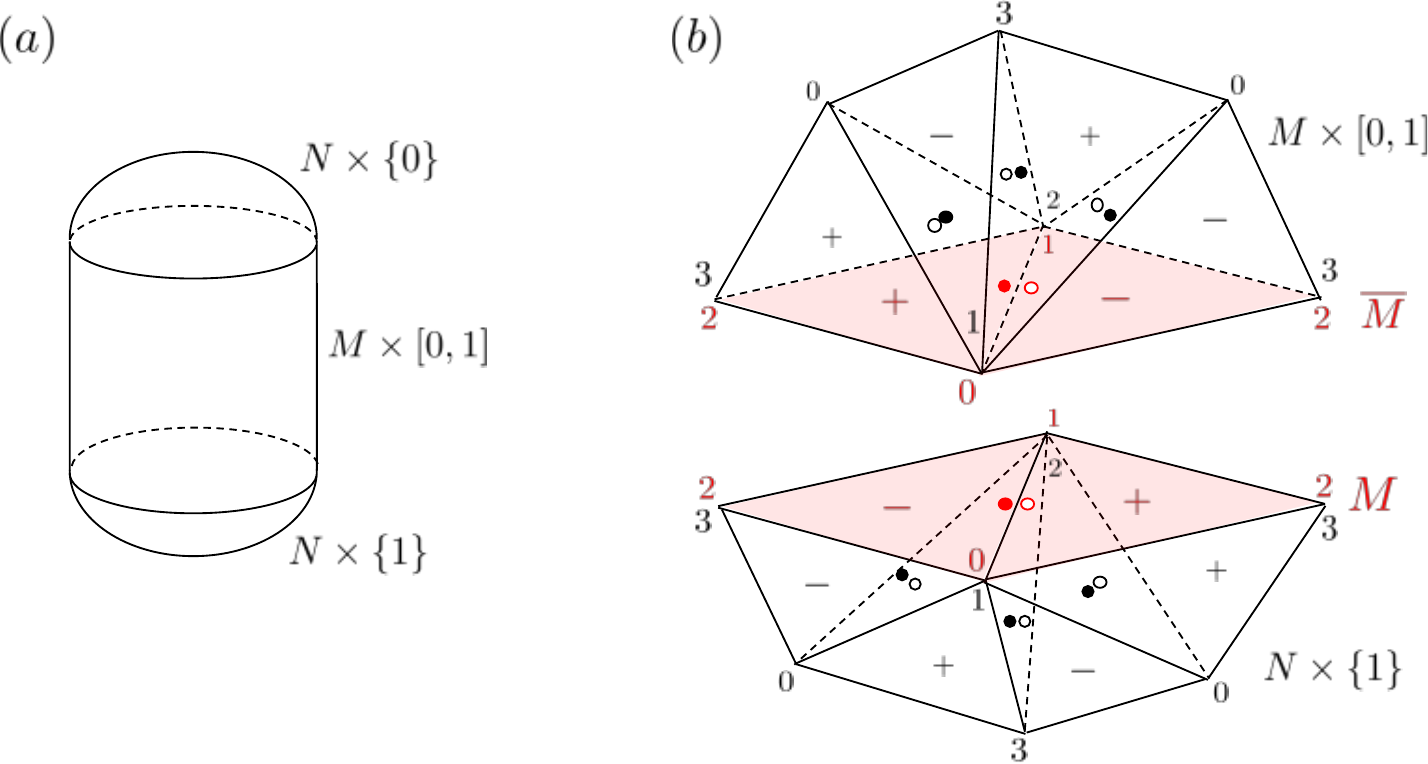}
\caption{$(a)$: An example of $K$ such that $\partial K= (N\times\{0\})\sqcup (M\times[0,1])\sqcup (N\times\{1\})$. $(b)$: Triangulation of $M\times[0,1]$ and $N\times\{1\}$ near the attaching region $M$, in the case of $d=2$. 
Note that $\theta$ (red dot) and $\overline{\theta}$ (white dot) are flipped from the original assignment rule on $M$ in the side of $N\times\{1\}$, which makes $\overline{\sigma}(N\times\{1\})=1$ when $\alpha(e)$ is nonzero on a single $(d-1)$-simplex on $M$.  
In contrast, we have $\sigma(M\times[0,1])=-1$ in such a situation.}
\label{fig:attach}
\end{figure}

To compare the value of the Gu-Wen integral on $N$ with different triangulations, we think of $K=N\times[0, 1]$, with the Gu-Wen integral on $\partial K= (N\times\{0\})\sqcup (M\times[0,1])\sqcup (N\times\{1\})$, see Fig.~\ref{fig:attach} $(a)$.
Suppose we have two triangulations and configurations of $(\alpha, \beta)$ we want to compare, on $N\times\{0\}$ and $N\times\{1\}$, respectively. 
We will compute the effect of re-triangulations by showing that
\begin{equation}
    \overline{\sigma}(N\times\{0\})\sigma(M\times[0,1])\overline{\sigma}(N\times\{1\})=(-1)^{\int_K\Sq^2(\beta)+w_2\cup\beta}.
    \label{eq:retriangle}
\end{equation}
Here we have extended $\beta\in Z^d(\partial K, \bZ_2)$ to $K$ on the right hand side of the above relation.

To see~\eqref{eq:retriangle}, we first observe the quadratic property of $\tilde{\sigma}(\alpha;\beta):=\overline{\sigma}(N\times\{0\})\sigma(M\times[0,1])\overline{\sigma}(N\times\{1\})$,
\begin{equation}
\tilde{\sigma}(\alpha;\beta)\tilde{\sigma}(\alpha';\beta')=\tilde{\sigma}(\alpha+\alpha';\beta+\beta')(-1)^{\int_{\partial K}\beta\cup_{d-1}\beta'},
\label{eq:Quadtilde}
\end{equation}
which can be seen by applying quadratic property of $\sigma$~\eqref{eq:Quadboundary} on $N\times\{0\}$, $M\times[0,1]$, $N\times\{1\}$.
Note that~\eqref{eq:Quadtilde} is satisfied for
\begin{equation}
    \tilde{\sigma}'(\alpha;\beta)=(-1)^{\int_{K}\Sq^2(\beta)},
\end{equation}
where we set $\Sq^2(\beta):=\beta\cup_{d-2}\beta+\delta\beta\cup_{d-1}\beta$. Thus, we can express $\tilde{\sigma}(\alpha;\beta)$ as $(-1)^{\int_K \Sq^2(\beta)}$ up to linear term,
\begin{equation}
   \tilde{\sigma}(\alpha;\beta)=(-1)^{\int_K \Sq^2(\beta)}(-1)^{\sum_{f\in K}\chi(f)\beta(f)}. 
\end{equation}
The linear term is fixed by computing $\tilde{\sigma}(\alpha;\beta)$ explicitly in the simplest case; 
$\beta=\delta\lambda$ on $\partial K$, and $\lambda(e)=1$ on a single $(d-1)$-simplex of $\partial K$, otherwise 0.
If we take a barycentric triangulation on $\partial K$, it is not hard to see that $\tilde{\sigma}(\alpha=0; \delta\lambda)=-1$ when $\lambda$ is nonzero away from the boundary of $N\times\{0\}$, $M\times[0,1]$, $N\times\{1\}$, by imitating the logic of Sec.~4.1.~in Gaiotto-Kapustin~\cite{Gaiotto:2015zta}.
In the case that $\lambda$ is nonzero on the boundary where $\lambda$ is identified as $\alpha$, we can show that 
$\overline{\sigma}(N\times\{0\})=1$, $\sigma(M\times[0,1])=-1$ (resp. $\overline{\sigma}(N\times\{1\})=1$, $\sigma(M\times[0,1])=-1$), when $\alpha(e)$ is nonzero on a single $(d-1)$-simplex on $M\times\{0\}$ (resp. $M\times\{1\}$). 
Thus, we have $\tilde{\sigma}(\lambda;\delta\lambda)=-1$, see Fig.~\ref{fig:attach} $(b)$.

Since the quadratic term $\Sq^2(\beta)$ vanishes for such $\beta$, the linear term is fixed as $(-1)^{\int_K w_2\cup\beta}$. 
Now we can write $\tilde{\sigma}(\alpha; \beta)=(-1)^{\int_K \Sq^2(\beta)+w_2\cup\beta}$, and
\begin{equation}
    \overline{\sigma}(N\times\{1\})=\overline{\sigma}(N\times\{0\})\cdot\sigma(M\times[0,1])(-1)^{\int_K \Sq^2(\beta)+w_2\cup\beta}.
\end{equation}
Next, we determine the form of quadratic function $\sigma(M\times[0,1])$ with the property~\eqref{eq:Quadboundary}. 
Note that~\eqref{eq:Quadboundary} is satisfied for
\begin{align}
    \sigma'(M\times[0,1])=(-1)^{\int_{M\times[0,1]} \Sq^2(\alpha)}.
\end{align}
Thus, we can express $\sigma(M\times[0,1])$ as $(-1)^{\int_{M\times[0,1]} \Sq^2(\alpha)}$ up to linear term,
\begin{equation}
   \sigma(M\times[0,1])=(-1)^{\int_{M\times[0,1]} \Sq^2(\alpha)}(-1)^{\sum_{e\in M\times[0,1]}\lambda(e)\alpha(e)}. 
\end{equation}
Note  that  $\alpha$ extends to $M\times[0,1]$ because $\alpha-\alpha'$ is assumed to be a coboundary.
The linear term is again fixed by computing $\sigma(M\times[0,1])$ explicitly in the simplest case; $\alpha(e)=1$ on a single $(d-1)$-simplex, otherwise 0. 
If we take a barycentric triangulation on $M\times[0,1]$, it is not hard to see that $\sigma(\alpha; \delta\alpha)=-1$ for such $\alpha$.
In the case that $\alpha$ is nonzero on boundary, we have to arrange in the computation that the orientation of $f_0$ is chosen to be identical to $t$.  

Thus, the linear term is fixed as $(-1)^{\sum_{e\in M\times[0,1]}\alpha(e)}$, where the sum runs over all $(d-1)$-simplices of a barycentric triangulation for $M\times[0,1]$. 
Therefore, we can write the linear term as $(-1)^{\int_{M\times[0,1]} w_2\cup\alpha}$ and
\begin{equation}
\sigma(M\times[0,1])=(-1)^{\int_{M\times[0,1]} \Sq^2(\alpha)+w_2\cup \alpha}. 
\end{equation}
Then, the variation of $\overline{\sigma}(\alpha;\beta)$ under re-triangulation and gauge transformation becomes
\begin{equation}
    (-1)^{\int_{M\times[0,1]}(\Sq^2(\alpha)+w_2\cup \alpha)+\int_{N\times[0,1]}(\Sq^2(\beta)+w_2\cup \beta)}.
\end{equation}
On the other hand, the variation of $(-1)^{\int_M\eta\cup\alpha+\int_N\eta\cup\beta}$ is given by
\begin{equation}
    (-1)^{\int_{M\times[0,1]}w_2\cup \alpha+\int_{N\times[0,1]}w_2\cup \beta}.
\end{equation}
Hence, the variation of the combined term $z[\eta; \alpha,\beta]=\overline{\sigma}(\alpha;\beta)(-1)^{\int_M\eta\cup\alpha+\int_N\eta\cup\beta}$ becomes
\begin{equation}
    (-1)^{\int_{M\times[0,1]}\Sq^2(\alpha) +\int_{N\times[0,1]}\Sq^2(\beta)}.
    \label{gwv}
\end{equation}

\subsection{Gapped boundary for the Gu-Wen phase}
After all these preparations, it is a simple matter to show that the boundary gauge theory \eqref{gwb} correctly couples to the bulk Gu-Wen SPT phase.
Indeed, the partition function of the coupled system has the action \begin{equation}
z[\eta;\alpha,\beta] (-1)^{-\int_M h^*x_d + \int_N g^*y_{d+1}} \label{combined}
\end{equation}
where we take $\alpha=h^*m_{d-1}$ and $\beta=g^*n_d$.
The first factor in \eqref{combined} has the variation \eqref{gwv},
whereas the second factor in \eqref{combined} has the variation 
\begin{equation}
(-1)^{\int_{M\times[0,1]} (h^*\delta x_d-g^*y_{d+1}) -\int_{N\times[0,1]} g^*\delta y_{d+1} }.
\end{equation}
These two variations cancel since we have  $\delta y_{d+1}=\Sq^2(n_d)$ and $y_{d+1}$ pulls back to $\Sq^2(m_{d-1})+\delta x_d$.
This is what we wanted to achieve.

\section*{Acknowledgements}
The authors thank Kazuya Yonekura for pointing out a serious flaw in a section originally contained in an earlier version of the paper, thereby saving the authors from a serious blunder on the arXiv.
RK is supported by Advanced Leading
Graduate Course for Photon Science (ALPS) of Japan Society for the Promotion of Science (JSPS).
KO is supported in part by the National Science Foundation grant PHY-1606531 and also  by the Paul Dirac fund.
YT is partially supported by JSPS KAKENHI Grant-in-Aid (Wakate-A), No.17H04837
and JSPS KAKENHI Grant-in-Aid (Kiban-S), No.16H06335,
and also by WPI Initiative, MEXT, Japan at IPMU, the University of Tokyo.

\bibliographystyle{ytphys}
\baselineskip=.95\baselineskip
\let\bbb\bibitem\def\bibitem{\itemsep3pt\bbb}
\bibliography{ref}

\providecommand{\href}[2]{#2}\begingroup\raggedright\begin{thebibliography}{10}

\bibitem{ABP}
D.~W. Anderson, E.~H. Brown, Jr., and F.~P. Peterson, {\slshape The structure
  of the {S}pin cobordism ring,} \href{http://dx.doi.org/10.2307/1970690}{{\em
  Ann. of Math. (2)} {\bfseries 86} (1967) 271--298}.

\bibitem{Sodemann:2016mib}
I.~Sodemann, I.~Kimchi, C.~Wang, and T.~Senthil, {\slshape {Composite Fermion
  Duality for Half-Filled Multicomponent Landau Levels},}
  \href{http://dx.doi.org/10.1103/PhysRevB.95.085135}{{\em Phys. Rev.}
  {\bfseries B95} (2017) 085135},
\href{http://arxiv.org/abs/1609.08616}{{ arXiv:1609.08616~[cond-mat.str-el]}}.

\bibitem{Wang:2017txt}
C.~Wang, A.~Nahum, M.~A. Metlitski, C.~Xu, and T.~Senthil, {\slshape
  {Deconfined Quantum Critical Points: Symmetries and Dualities},}
  \href{http://dx.doi.org/10.1103/PhysRevX.7.031051}{{\em Phys. Rev.}
  {\bfseries X7} (2017) 031051},
\href{http://arxiv.org/abs/1703.02426}{{ arXiv:1703.02426~[cond-mat.str-el]}}.

\bibitem{Garcia-Etxebarria:2017crf}
I.~García-Etxebarria, H.~Hayashi, K.~Ohmori, Y.~Tachikawa, and K.~Yonekura,
  {\slshape {8d gauge anomalies and the topological Green-Schwarz mechanism},}
  \href{http://dx.doi.org/10.1007/JHEP11(2017)177}{{\em JHEP} {\bfseries 11}
  (2017) 177},
\href{http://arxiv.org/abs/1710.04218}{{ arXiv:1710.04218~[hep-th]}}.

\bibitem{CordovaOhmoriToAppear}
C.~C{\'o}rdova and K.~Ohmori. To appear.

\bibitem{Seiberg:2016rsg}
N.~Seiberg and E.~Witten, {\slshape {Gapped Boundary Phases of Topological
  Insulators via Weak Coupling},}
  \href{http://dx.doi.org/10.1093/ptep/ptw083}{{\em PTEP} {\bfseries 2016}
  (2016) 12C101},
\href{http://arxiv.org/abs/1602.04251}{{ arXiv:1602.04251~[cond-mat.str-el]}}.

\bibitem{Witten:2016cio}
E.~Witten, {\slshape {The "Parity" Anomaly on an Unorientable Manifold},}
  \href{http://dx.doi.org/10.1103/PhysRevB.94.195150}{{\em Phys. Rev.}
  {\bfseries B94} (2016) 195150},
\href{http://arxiv.org/abs/1605.02391}{{ arXiv:1605.02391~[hep-th]}}.

\bibitem{Wang:2017loc}
J.~Wang, X.-G. Wen, and E.~Witten, {\slshape {Symmetric Gapped Interfaces of
  SPT and SET States: Systematic Constructions},}
  \href{http://dx.doi.org/10.1103/PhysRevX.8.031048}{{\em Phys. Rev.}
  {\bfseries X8} (2018) 031048},
\href{http://arxiv.org/abs/1705.06728}{{ arXiv:1705.06728~[cond-mat.str-el]}}.

\bibitem{Tachikawa:2017gyf}
Y.~Tachikawa, {\slshape {On Gauging Finite Subgroups},}
\href{http://arxiv.org/abs/1712.09542}{{ arXiv:1712.09542~[hep-th]}}.

\bibitem{Hsieh:2018ifc}
C.-T. Hsieh, {\slshape {Discrete Gauge Anomalies Revisited},}
\href{http://arxiv.org/abs/1808.02881}{{ arXiv:1808.02881~[hep-th]}}.

\bibitem{Guo:2018vij}
M.~Guo, K.~Ohmori, P.~Putrov, Z.~Wan, and J.~Wang, {\slshape {Fermionic
  Finite-Group Gauge Theories and Interacting Symmetric/Crystalline Orders via
  Cobordisms},}
\href{http://arxiv.org/abs/1812.11959}{{ arXiv:1812.11959~[hep-th]}}.

\bibitem{WangOhmori2018}
J.~Wang, K.~Ohmori, P.~Putrov, Y.~Zheng, Z.~Wan, M.~Guo, H.~Lin, P.~Gao, and
  S.~T. Yau, {\slshape {Tunneling topological vacua via extended operators:
  (Spin-)TQFT spectra and boundary deconfinement in various dimensions},}
  \href{http://dx.doi.org/10.1093/ptep/pty051}{{\em Progress of Theoretical and
  Experimental Physics} {\bfseries 2018} (2018) 0--59},
  \href{http://arxiv.org/abs/1801.05416v3}{{ arXiv:1801.05416v3}}.

\bibitem{ThorngrenKeyserlingk2015}
R.~Thorngren and C.~von Keyserlingk, {\slshape {Higher SPT's and a
  generalization of anomaly in-flow},} \href{http://arxiv.org/abs/1511.02929}{{
  arXiv:1511.02929}}.

\bibitem{Kobayashi2019}
R.~Kobayashi and K.~Shiozaki, {\slshape {Anomaly indicator of rotation symmetry
  in (3+1)D topological order},} \href{http://arxiv.org/abs/1901.06195}{{
  arXiv:1901.06195}}.

\bibitem{Kapustin2014anomalous}
A.~Kapustin and R.~Thorngren, {\slshape {Anomalous discrete symmetries in three
  dimensions and group cohomology},}
  \href{http://dx.doi.org/10.1103/PhysRevLett.112.231602}{{\em Physical Review
  Letters} {\bfseries 112} (2014) 1--31},
  \href{http://arxiv.org/abs/1404.3230v2}{{ arXiv:1404.3230v2}}.

\bibitem{Fidkowski2014}
L.~Fidkowski, X.~Chen, and A.~Vishwanath, {\slshape {Non-Abelian topological
  order on the surface of a 3d topological superconductor from an exactly
  solved model},} \href{http://dx.doi.org/10.1103/PhysRevX.3.041016}{{\em
  Physical Review X} {\bfseries 3} (2014) 041016},
  \href{http://arxiv.org/abs/1305.5851v4}{{ arXiv:1305.5851v4}}.

\bibitem{Cheng2017exactly}
M.~Cheng, Z.~C. Gu, S.~Jiang, and Y.~Qi, {\slshape {Exactly solvable models for
  symmetry-enriched topological phases},}
  \href{http://dx.doi.org/10.1103/PhysRevB.96.115107}{{\em Physical Review B}
  {\bfseries 96} (2017) 47--49}, \href{http://arxiv.org/abs/1606.08482v3}{{
  arXiv:1606.08482v3}}.

\bibitem{Chen2016symmetry}
X.~Chen and M.~Hermele, {\slshape {Symmetry fractionalization and anomaly
  detection in three-dimensional topological phases},}
  \href{http://dx.doi.org/10.1103/PhysRevB.94.195120}{{\em Physical Review B}
  {\bfseries 94} (2016) 1--19}, \href{http://arxiv.org/abs/1602.00187v1}{{
  arXiv:1602.00187v1}}.

\bibitem{Chen2015anomalous}
X.~Chen, F.~J. Burnell, A.~Vishwanath, and L.~Fidkowski, {\slshape {Anomalous
  symmetry fractionalization and surface topological order},}
  \href{http://dx.doi.org/10.1103/PhysRevX.5.041013}{{\em Physical Review X}
  {\bfseries 5} (2015) 1--21}, \href{http://arxiv.org/abs/1403.6491v2}{{
  arXiv:1403.6491v2}}.

\bibitem{Barkeshli2016}
M.~Barkeshli, P.~Bonderson, C.-M. Jian, M.~Cheng, and K.~Walker, {\slshape
  {Reflection and time reversal symmetry enriched topological phases of matter:
  path integrals, non-orientable manifolds, and anomalies},}
  \href{http://arxiv.org/abs/1612.07792}{{ arXiv:1612.07792}}.

\bibitem{Chen:2011pg}
X.~Chen, Z.-C. Gu, Z.-X. Liu, and X.-G. Wen, {\slshape {Symmetry Protected
  Topological Orders and the Group Cohomology of Their Symmetry Group},}
  \href{http://dx.doi.org/10.1103/PhysRevB.87.155114}{{\em Phys. Rev.}
  {\bfseries B87} (2013) 155114},
\href{http://arxiv.org/abs/1106.4772}{{ arXiv:1106.4772~[cond-mat.str-el]}}.

\bibitem{Kapustin:2014tfa}
A.~Kapustin, {\slshape {Symmetry Protected Topological Phases, Anomalies, and
  Cobordisms: Beyond Group Cohomology},}
\href{http://arxiv.org/abs/1403.1467}{{ arXiv:1403.1467~[cond-mat.str-el]}}.

\bibitem{Kapustin:2014dxa}
A.~Kapustin, R.~Thorngren, A.~Turzillo, and Z.~Wang, {\slshape {Fermionic
  Symmetry Protected Topological Phases and Cobordisms},}
  \href{http://dx.doi.org/10.1007/JHEP12(2015)052}{{\em JHEP} {\bfseries 12}
  (2015) 052},
\href{http://arxiv.org/abs/1406.7329}{{ arXiv:1406.7329~[cond-mat.str-el]}}.

\bibitem{Freed:2016rqq}
D.~S. Freed and M.~J. Hopkins, {\slshape {Reflection Positivity and Invertible
  Topological Phases},}
\href{http://arxiv.org/abs/1604.06527}{{ arXiv:1604.06527~[hep-th]}}.

\bibitem{Yonekura:2018ufj}
K.~Yonekura, {\slshape {On the Cobordism Classification of Symmetry Protected
  Topological Phases},}
\href{http://arxiv.org/abs/1803.10796}{{ arXiv:1803.10796~[hep-th]}}.

\bibitem{Gu:2012ib}
Z.-C. Gu and X.-G. Wen, {\slshape {Symmetry-protected topological orders for
  interacting fermions: Fermionic topological nonlinear $\sigma$ models and a
  special group supercohomology theory},}
  \href{http://dx.doi.org/10.1103/PhysRevB.90.115141}{{\em Phys. Rev.}
  {\bfseries B90} (2014) 115141},
\href{http://arxiv.org/abs/1201.2648}{{ arXiv:1201.2648~[cond-mat.str-el]}}.

\bibitem{Gaiotto:2015zta}
D.~Gaiotto and A.~Kapustin, {\slshape {Spin TQFTs and Fermionic Phases of
  Matter},} \href{http://dx.doi.org/10.1142/S0217751X16450445}{{\em Int. J.
  Mod. Phys.} {\bfseries A31} (2016) 1645044},
\href{http://arxiv.org/abs/1505.05856}{{ arXiv:1505.05856~[cond-mat.str-el]}}.

\bibitem{Gaiotto:2014kfa}
D.~Gaiotto, A.~Kapustin, N.~Seiberg, and B.~Willett, {\slshape {Generalized
  Global Symmetries},} \href{http://dx.doi.org/10.1007/JHEP02(2015)172}{{\em
  JHEP} {\bfseries 02} (2015) 172},
\href{http://arxiv.org/abs/1412.5148}{{ arXiv:1412.5148~[hep-th]}}.

\bibitem{Hsin:2019fhf}
P.-S. Hsin and A.~Turzillo, {\slshape {Symmetry-Enriched Quantum Spin Liquids
  in $(3+1)d$},}
\href{http://arxiv.org/abs/1904.11550}{{ arXiv:1904.11550~[cond-mat.str-el]}}.

\bibitem{Freed:2018cec}
D.~S. Freed and C.~Teleman, {\slshape {Topological Dualities in the Ising
  Model},}
\href{http://arxiv.org/abs/1806.00008}{{ arXiv:1806.00008~[math.AT]}}.

\bibitem{Thorngren:2014pza}
R.~Thorngren, {\slshape {Framed Wilson Operators, Fermionic Strings, and
  Gravitational Anomaly in 4D},}
  \href{http://dx.doi.org/10.1007/JHEP02(2015)152}{{\em JHEP} {\bfseries 02}
  (2015) 152},
\href{http://arxiv.org/abs/1404.4385}{{ arXiv:1404.4385~[hep-th]}}.

\bibitem{BrownPeterson}
E.~H. Brown, Jr. and F.~P. Peterson, {\slshape Relations among characteristic
  classes. {I},} \href{http://dx.doi.org/10.1016/0040-9383(64)90004-7}{{\em
  Topology} {\bfseries 3} (1964) 39--52}.

\bibitem{Conner}
P.~E. Conner, \href{http://dx.doi.org/10.1007/BFb0063217}{{\em Differentiable
  periodic maps}}, vol.~738 of {\em Lecture Notes in Mathematics}.
\newblock Springer, Berlin, second~ed., 1979.

\bibitem{PengelleyWilliams}
D.~J. Pengelley and F.~Williams, {\slshape Global structure of the mod two
  symmetric algebra, {$H^*(BO;\mathbb{F}_2)$}, over the {S}teenrod algebra,}
  \href{http://dx.doi.org/10.2140/agt.2003.3.1119}{{\em Algebr. Geom. Topol.}
  {\bfseries 3} (2003) 1119--1138},
  \href{http://arxiv.org/abs/math.AT/0312220}{{ arXiv:math.AT/0312220}}.

\bibitem{MS}
J.~W. Milnor and J.~D. Stasheff, {\em Characteristic classes}.
\newblock Princeton University Press, Princeton, N. J.; University of Tokyo
  Press, Tokyo, 1974.
\newblock Annals of Mathematics Studies, No. 76.

\bibitem{ManifoldAtlasWu}
{Karlheinz Knapp}, {\slshape Wu class,}
  \url{http://www.map.mpim-bonn.mpg.de/Wu_class}.

\bibitem{BFquadratic}
G.~Brumfiel and J.~Morgan, {\slshape {Quadratic Functions of Cocycles and Pin
  Structures},} \href{http://arxiv.org/abs/1808.10484}{{
  arXiv:1808.10484~[math.AT]}}.

\bibitem{HalperinToledo}
S.~Halperin and D.~Toledo, {\slshape Stiefel-{W}hitney homology classes,}
  \href{http://dx.doi.org/10.2307/1970823}{{\em Ann. of Math. (2)} {\bfseries
  96} (1972) 511--525}.

\bibitem{BlantonMcCrory}
J.~D. Blanton and C.~McCrory, {\slshape An axiomatic proof of {S}tiefel's
  conjecture,} \href{http://dx.doi.org/10.2307/2042195}{{\em Proc. Amer. Math.
  Soc.} {\bfseries 77} (1979) 409--414}.

\end{thebibliography}\endgroup

\end{document}